\documentclass[conference]{IEEEtran}

\usepackage{cite}
\usepackage{amsmath,amssymb}
\usepackage{graphicx}
\usepackage{url}
\usepackage{color}
\usepackage{algorithm}
\usepackage{algpseudocode}
\usepackage{booktabs}
\usepackage{multirow}
\usepackage{array}
\usepackage{stfloats}
\usepackage{longtable}
\usepackage[hidelinks]{hyperref}
\usepackage{xurl} 
\begin{document}

\title{Quality-Driven Agentic Reasoning for LLM-Assisted Software Design: Questions-of-Thoughts (QoT) as a Time-Series Self-QA Chain}

\author{
    \IEEEauthorblockN{Yen-Ku Liu\IEEEauthorrefmark{1}, Yun-Cheng Tsai\IEEEauthorrefmark{1}}
    \IEEEauthorblockA{\IEEEauthorrefmark{1}Department of Technology Application and Human Resource Development,\\ National Taiwan Normal University, Taipei, Taiwan\\ Email: pecu@ntnu.edu.tw}
}

\maketitle

\begin{abstract}
Recent advances in large language models (LLMs) have accelerated AI-assisted software development, yet practical deployment remains constrained by incomplete implementations, weak modularization, and inconsistent security practices. We introduce Questions-of-Thoughts (QoT), a quality-driven inference-time scaffold that turns a user goal into (i) an ordered sequence of engineering steps and (ii) stepwise self-questioning to verify constraints and reduce omission errors, while maintaining a lightweight reasoning record that stabilizes subsequent design decisions.

We evaluate QoT across three representative backend engineering domains: API Design, Data Communication, and File Systems. Each task requires multi-module decomposition and exposes standard failure modes in LLM-generated systems. To enable data-driven comparison, we score generated artifacts using an ISO/IEC-inspired quality rubric that measures Scalability, Completeness, Modularity, and Security. We report domain-wise gains as the change in total quality score, defined as the QoT score minus the NoQoT score. Results show capacity-dependent improvements: QoT yields consistent quality improvements for larger models and more complex domains, while smaller models may exhibit trade-offs under tight context and planning budgets.

We release an open artifact with prompts, scoring guidelines, raw generations, and scripts that reproduce the reported tables and figures to support applied AI and data analytics research.

\end{abstract}


\section{Introduction}
Large Language Models (LLMs) have rapidly evolved from ``autocomplete'' assistants into \emph{agentic} systems that write code, invoke tools, and iteratively refine their outputs.
LLM agents are driving this shift in software engineering by carrying out end-to-end workflows, including diagnosing issues, patching code, running tests, and producing documentation, under real-world constraints.
However, practical adoption still faces three recurring pain points: (i) \textbf{unreliable correctness} under distribution shift and long-horizon tasks, (ii) \textbf{limited evidence explaining why a reviewer or practitioner should trust a solution}, and (iii) \textbf{unstable evaluation} caused by contamination and benchmark overfitting.
This reliability gap is especially salient for applied AI pipelines and analytics-driven systems that must meet operational constraints, auditability requirements, and risk-based compliance.

Recent work has responded by building more realistic and robust evaluation suites.
LiveBench emphasizes contamination-limited, frequently updated tasks with objective scoring to keep evaluation ``fresh'' as models improve \cite{white2025livebench}.
For software engineering in particular, SWE-Bench Pro extends SWE-Bench toward more complex, enterprise-like, multi-file edits that better approximate professional work \cite{deng2025swebenchpro}.
Beyond bug-fixing, security-oriented benchmarks such as SEC-bench highlight that \emph{functional correctness alone} is insufficient: agents must also avoid introducing vulnerabilities. They must reason under adversarial conditions \cite{lee2025secbench}.
At the agent level, the agent company evaluates LLM agents in a simulated workplace where browsing, coding, running programs, and communication are required to complete consequential tasks \cite{xu2024theagentcompany}.
Complementing these benchmarks, a growing body of survey work systematizes evaluation dimensions (task design, environment fidelity, metrics, and failure modes) and argues for standardized, reproducible agent assessments \cite{mohammadi2025evalsurvey}.

In parallel, the agentic methods landscape has expanded from single-shot generation to \emph{search and verification at inference time}.
Agent-guided tree search frameworks explore alternative solution strategies and refine candidate solutions in a structured way, improving both coverage and the final selection for code generation \cite{li2025codetree}.
Execution-centric agents further close the loop by automatically configuring and running project test suites, producing actionable feedback that mimics how developers validate changes \cite{bouzenia2025executionagent}.
More generally, verification-augmented refinement pipelines split complex objectives into checkable constraints and iterate based on tool-grounded signals \cite{zhang2025dvr}.
Finally, test-time scaling methods increase inference compute via parallel generation, sequential refinement, and execution-grounded discrimination, substantially improving code generation performance across model families \cite{li2025sstar}.

Despite this progress, a central gap remains: many existing agent pipelines primarily optimize pass rates without explicitly modeling software quality attributes (e.g., reliability, maintainability, and clarity) that matter to downstream users and organizations.
In practice, developers often need a system that does not merely produce a passing patch, but also provides \emph{transparent, reusable reasoning artifacts} that support review, debugging, and governance.
This challenge aligns directly with ongoing priorities in AI and data analytics across foundations, frontiers, and responsible deployment, especially explainability, trustworthy pipelines, and agentic systems for software engineering.

To address these needs, we propose \textbf{Question-of-Thoughts (QoT)}, a quality-driven, question-centric reasoning framework for LLM agents in code-related tasks.
QoT organizes reasoning as an explicit sequence of targeted questions that (1) elicit critical constraints early, (2) guide tool use and intermediate verification, and (3) record a structured trace that can be audited and reused.
We instantiate QoT in an LLM agent that generates and refines code solutions while tracking quality-oriented criteria, and we evaluate its impact on both functional and quality-focused outcomes.
To strengthen realism, we also incorporate deployment-oriented trustworthiness checks inspired by recent work on validating agentic RAG systems under practical constraints \cite{kale2025monitoring}.

\vspace{2pt}
\noindent\textbf{Contributions.} This work sits at the intersection of LLM-assisted software engineering, agentic systems, and trustworthy AI for deployable analytics and engineering pipelines. We contribute:
\begin{itemize}
  \item a structured agentic reasoning protocol (QoT) that couples sequential planning with stepwise self-QA and constraint tracking;
  \item an evaluation benchmark spanning three practical backend domains (API Design, Data Communication, and File Systems);
  \item a quality-driven, data-centric evaluation rubric grounded in ISO/IEC notions to enable reproducible cross-model comparison; and
  \item an open artifact with prompts, scoring guidelines, and scripts to reproduce key results.
\end{itemize}

In summary, this paper makes three contributions:
\begin{enumerate}
    \item We introduce the QoT framework for quality-aware agent reasoning in software tasks.
    \item We implement an agentic workflow that integrates QoT with verification and refinement signals.
    \item We provide an empirical study that analyzes when and why QoT improves outcome quality, with implications for reliable AI and data analytics systems.
\end{enumerate}

\section{Literature Review}
\subsection{LLMs for Code Generation and Software Engineering}
Large language models have demonstrated strong performance in code completion and program synthesis, supported by scaling and large-scale pretraining \cite{brown2022language}. Code-oriented systems such as Codex and AlphaCode show that LLMs can generate functional programs and assist with development tasks \cite{chen2021evaluating}. Nevertheless, moving from ``working snippets'' to deployable systems requires meeting engineering constraints, including modular design, error handling, validation, and secure defaults, which remain challenging under one-shot generation.

\subsection{Inference-Time Structured Reasoning and Agentic Workflows}
Inference-time reasoning methods aim to improve output quality without additional training. CoT prompting \cite{wei2022chain} improves reasoning by encouraging explicit intermediate steps, while ToT \cite{yao2023tree} generalizes this idea by exploring multiple branches and selecting promising candidates. These approaches enhance logical consistency, but software engineering problems typically require dependency-aware planning across modules and iterative verification of constraints (e.g., authentication, authorization, data validation, and failure modes). Recent self-edit and self-refinement paradigms further motivate the integration of verification loops into generation pipelines \cite{Yueke_Zhang_2025}.

Unlike typical self-refinement approaches that iteratively improve generated code after an initial solution is produced, QoT emphasizes structured constraint elicitation before and during generation. By organizing reasoning as a sequence of targeted questions aligned with software quality criteria, QoT encourages earlier identification of design constraints and potential omissions, which helps stabilize downstream reasoning and refinement.

\subsection{Reliability and Quality Criteria for Generated Code}
A recurring issue in LLM-generated code is that evaluation often focuses on functional plausibility with small examples. At the same time, practical use depends on non-functional properties such as scalability, maintainability, and security. Studies show that LLM-generated solutions often omit corner cases and system-level structure, leading to brittle designs \cite{mohammadi2025evalsurvey}. Software quality standards (e.g., ISO/IEC 9126 and ISO/IEC 25010) provide operational characteristics for assessing artifacts, and engineering practice often emphasizes completeness, modularity, scalability, and risk controls for production deployment.
\subsection{Gap and Motivation}
Existing reasoning paradigms do not explicitly integrate
\begin{enumerate}
    \item sequential dependency tracking,
    \item systematic self-verification aligned with software quality criteria, and
    \item accumulation of intermediate decisions into a reusable reasoning state for subsequent refinement.
\end{enumerate}

These limitations motivate QoT, which combines a sequential process chain with a self-QA verification layer and an evolving reasoning knowledge base to improve completeness, modularity, and security in LLM-assisted software engineering.

\section{Architecture}

The \textit{Questions-of-Thoughts} (QoT) algorithm is an inference-time, agentic reasoning pipeline designed to improve \textbf{quality} and \textbf{reliability} in LLM-assisted software engineering. QoT consists of three components: the \textbf{Sequential Process Chain} for dependency-aware decomposition, the \textbf{Question--Answer (self-QA) Chain} for stepwise clarification and verification, and the \textbf{Reasoning Knowledge Base} that accumulates intermediate decisions and constraints. These components work together to guide generation toward deployable system designs rather than isolated code fragments.

\begin{figure*}[!t]
    \centering
    \includegraphics[width=0.75\linewidth]{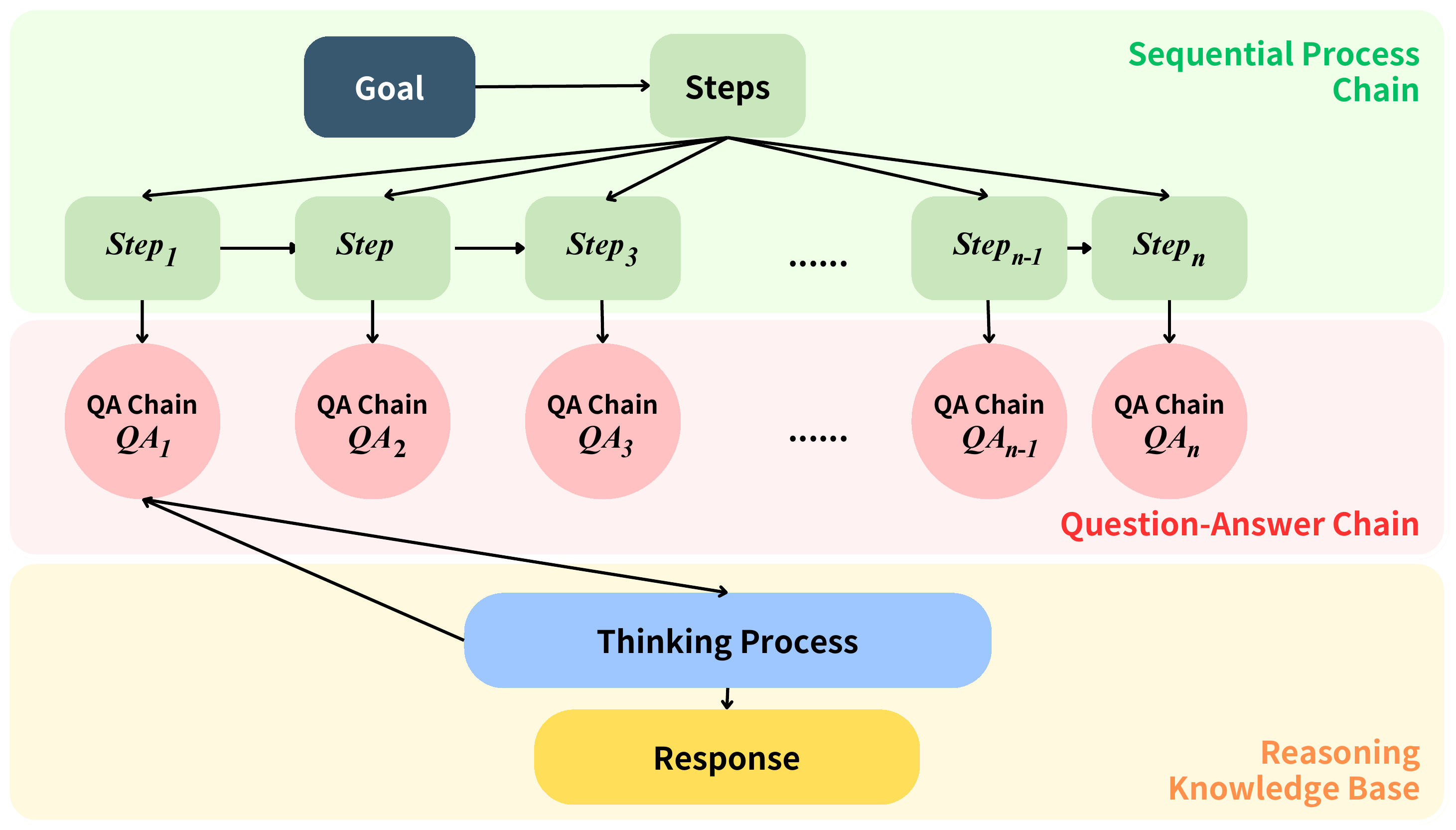}
    \caption{Overall Structure of the Questions-of-Thoughts (QoT) Algorithm}
    \label{fig:architecture}
\end{figure*}

\subsection{Sequential Process Chain}

The \textbf{Sequential Process Chain} module systematically breaks a high-level problem into a sequence of ordered steps. This decomposition allows the model to process information progressively, maintaining logical consistency while avoiding redundancy or incomplete responses.

When the system receives an objective  \( G \), it first decomposes the objective into an ordered sequence of steps:

\[
\text{Steps} = \{ S_1, S_2, \dots, S_n \}
\]

Each step \( S_i \) represents a structured sub-task that forms a logical progression centering on problem definition. This decomposition ensures that the reasoning path remains structured and allows the system to handle each sub-problem efficiently.

\subsection{Question-Answer Chain}

The \textbf{Question-Answer Chain} extends the reasoning process by employing a structured self-questioning mechanism at each step. Inspired by the Socratic method, this module allows the model to refine its responses through iterative sequential questioning.

For each step \( S_i \), a corresponding set of self-generated questions is formed:

\[
\text{QA}_i = \{ Q_{i,1}, Q_{i,2}, \dots, Q_{i,m_i} \}
\]

\begin{figure*}[!t]
    \centering
    \includegraphics[width=0.75\linewidth]{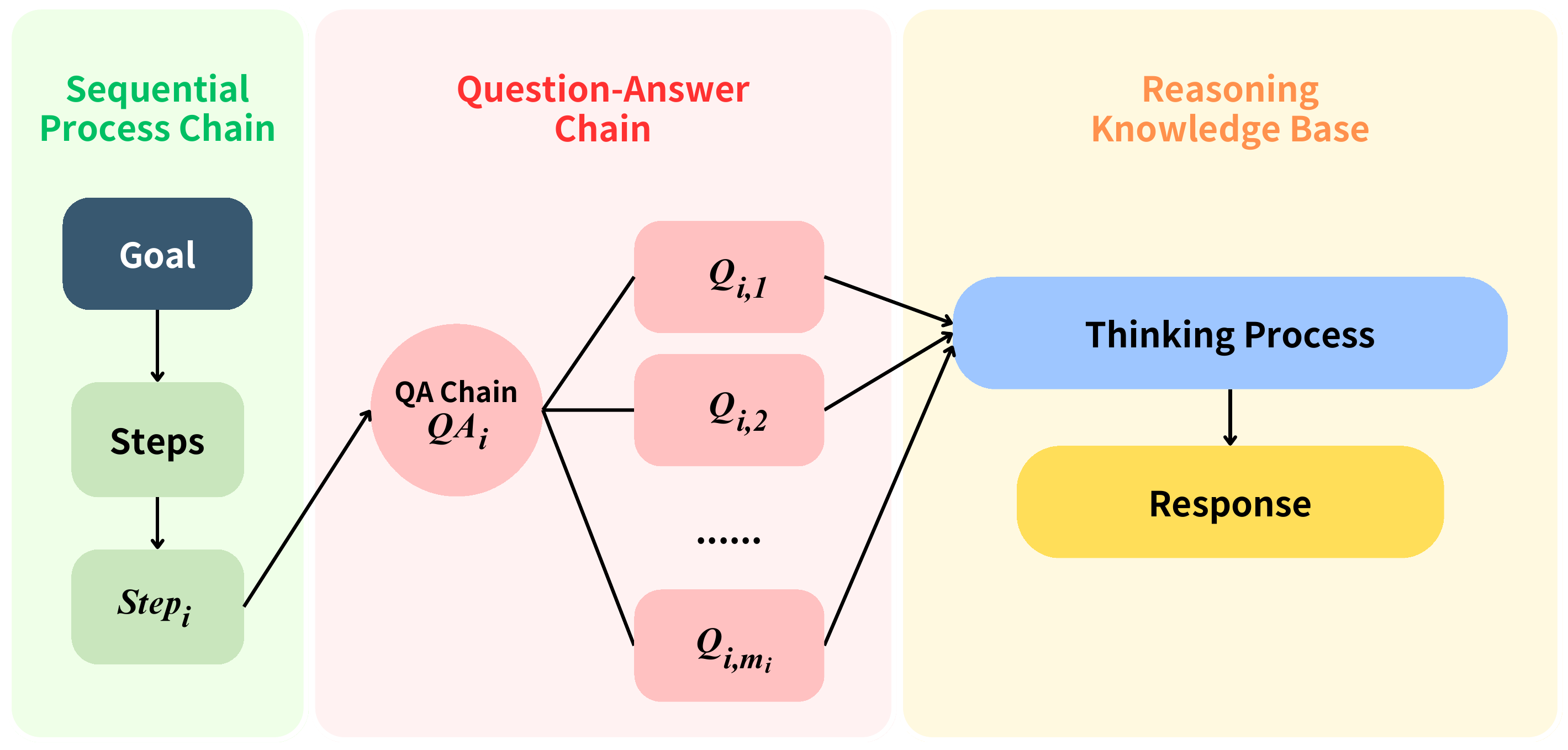}
    \caption{Sequential QA Chain Construction in QoT}
    \label{fig:qa_chain}
\end{figure*}

Where each question \( Q_{i,j} \) seeks to clarify, expand, or verify information related to \( S_i \). This iterative approach ensures the generated code or response is complete, logically sound, and aligned with engineering best practices. Figure~\ref{fig:qa_chain} illustrates the structure of the question-answer mechanism.

\subsection{Reasoning Knowledge Base}

The \textbf{Reasoning Knowledge Base} serves as the decision-making and memory module of QoT. It accumulates the reasoning process, tracks dependencies, and uses prior insights to refine subsequent responses.

The knowledge base maintains two core components:
\begin{itemize}
    \item \textbf{Thinking Process (TP)}: A dynamic storage of intermediate reasoning steps and accumulated insights.
    \item \textbf{Response (R)}: A progressively updated solution that integrates knowledge from TP.
\end{itemize}

The iterative nature of this process ensures that all reasoning steps contribute meaningfully to the final response. Algorithm~\ref{alg:qot} outlines the structured execution of this method.

\begin{algorithm}[t]
\caption{QoT-Reasoning Algorithm}
\label{alg:qot}
\begin{algorithmic}[1]
\Require Given objective \( G \)
\State \textbf{Input:} User-defined problem statement \( G \)
\State \textbf{Output:} Optimized final response \( R \)
\State Initialize \( TP \gets \emptyset \) \Comment{Thinking Process Storage}
\State Initialize \( R \gets \emptyset \) \Comment{Final Response}

\State \textbf{Step 1: Generate Sequential Steps}
\State \( \text{Steps} \gets \text{GenerateSteps}(G) \)
\If{\text{Error in Steps}}
    \State \textbf{Return} Error Message
\EndIf

\State \textbf{Step 2: Generate Question-Answer Chain}
\For{each step \( S_i \in \text{Steps} \)}
    \State \( \text{QA}_i \gets \text{GenerateQuestions}(S_i) \)
    \If{\text{Error in Questions}}
        \State \textbf{Return} Error Message
    \EndIf

    \For{each question \( Q_{i,j} \in QA_i \)}
        \State \( \text{Answer}_{i,j} \gets \text{GenerateAnswer}(Q_{i,j}) \)
        \If{\text{Error in Answer}}
            \State \textbf{Return} Error Message
        \EndIf
        \State \( TP \gets TP \cup \text{Answer}_{i,j} \) \Comment{Accumulate reasoning steps}
    \EndFor
\EndFor

\State \textbf{Step 3: Update Final Response}
\State \( R \gets \text{UpdateFinalResponse}(TP) \)

\State \textbf{Return} \( R \)
\end{algorithmic}
\end{algorithm}

In the current implementation, if an error occurs during step generation or question answering, the system terminates the current reasoning branch and returns a structured error message. In practical deployments, this mechanism can be extended with retry strategies or simplified fallback prompts to improve robustness in longer agentic workflows.

By employing a sequential process chain, a structured self-QA mechanism, and an evolving reasoning knowledge base, QoT improves completeness, reliability, and modularity for LLM-assisted software development. Our open-source implementation is available at \url{https://github.com/knyliu/questions-of-thoughts}, enabling reproducible evaluation and extensions.

\section{Implementation and Results}
To evaluate the effectiveness of the \textit{Questions-of-Thoughts} (QoT) algorithm for improving LLM-assisted software design, we conducted controlled experiments across three domains that capture common patterns in applied AI and enterprise systems: \textbf{API design}, \textbf{data communication}, and \textbf{file systems}. We focus on inference-time quality improvement: we leave the underlying base models intact and apply QoT as an external reasoning protocol.

\subsection{Evaluation Setup}
We evaluate QoT on three representative backend engineering domains: \emph{API Design}, \emph{Data Communication}, and \emph{File Systems}.
Across all domains, each benchmark task requires multi-module decomposition (e.g., user management, core logic, and persistence/notification behaviors) and explicitly emphasizes practical engineering concerns such as input validation, defensive error handling, access control, and failure handling.
Table~\ref{tab:tasks} summarizes the three tasks and their key modules.
Full prompt text and model outputs are released in our artifact to support reproducibility.

\begin{table*}[!t]
\centering
\normalsize
\setlength{\tabcolsep}{5pt}
\renewcommand{\arraystretch}{1.15}
\caption{Benchmark domains and task summaries used in our evaluation. The artifact provides full prompt text for reproducibility.}
\label{tab:tasks}
\begin{tabular}{c | p{0.3\linewidth} | p{0.2\linewidth} | p{0.2\linewidth}}
\toprule
\textbf{Domain} & \textbf{Task summary} & \textbf{Key aspects} & \textbf{Modules} \\
\midrule
API Design &
Room booking backend with registration/login, room CRUD, reservation lifecycle, and notifications. &
Validation, access control, defensive errors &
User, Room, Booking, Notify \\
\midrule
Data Communication &
Real-time messaging API with send/receive, timestamps, and history query/filtering. &
Schema checks, retries, basic observability &
User, Message, History \\
\midrule
File Systems &
File management API with upload/download/delete, metadata persistence, and per-user listing. &
Permissions, metadata integrity, failure-handling &
User, File, Metadata \\
\bottomrule
\end{tabular}
\end{table*}

\subsection{Evaluation Metrics}
Deployable software quality requires more than functional correctness. We evaluate each generated output using a rubric grounded in ISO/IEC 9126 and ISO/IEC 25010 quality characteristics and informed by the FURPS perspective. For every response, we score four dimensions: \textbf{Scalability} (Sc), \textbf{Completeness} (Co), \textbf{Modularity} (Mo), and \textbf{Security} (Se). We assign scores on a four-level scale from 1 to 4, where four indicates the response fully meets the criterion and one indicates it does not. We then aggregate the scores across the three domains.

The evaluation is conducted via an automated LLM-as-a-judge framework, utilizing a computationally capable LLM with a long context window (Gemini 3.1 Pro Preview) acting as the evaluator. This automated setup avoids human scoring variability while ensuring that the same evaluation criteria are applied consistently across all experimental conditions. To prevent methodological bias, the evaluation process is strictly double-blinded; the origin of each generated codebase is anonymized and presented to the evaluator simply as \textit{Method\_A}, \textit{Method\_B}, and \textit{Method\_C}.

To further stabilize scoring behavior across repeated runs, we introduce an $N-1$ overlapping context strategy. During the evaluation of round $N$, the evaluation context includes the raw source codes and the raw quantitative scores from the immediately preceding round ($N-1$). This overlapping evaluation context helps maintain consistent grading standards for the LLM judge and reduces potential evaluation drift across independent evaluation rounds.

\begin{table*}
\centering
\normalsize
\setlength{\tabcolsep}{6pt}
\renewcommand{\arraystretch}{1.15}
\caption{Quality score improvements across domains for QoT compared with two baselines. Values are reported as mean $\pm$ standard deviation across runs. Positive values indicate improvements.}
\label{tab:qot_gain}
\textbf{(a) QoT $-$ NoQoT}

\begin{tabular}{l c rrr}
\toprule
\textbf{Model} & \textbf{Setting} & \textbf{$\Delta$API} & \textbf{$\Delta$Comm} & \textbf{$\Delta$FS} \\
\midrule
llama3.1\_8b  & QoT--NoQoT & $1.40 \pm 2.07$ & $2.60 \pm 3.97$ & $1.00 \pm 2.55$ \\
llama3.1\_70b & QoT--NoQoT & $3.40 \pm 1.34$ & $5.40 \pm 1.67$ & $-2.80 \pm 1.10$ \\
llama3.2\_3b  & QoT--NoQoT & $4.60 \pm 1.67$ & $5.40 \pm 1.67$ & $3.60 \pm 3.78$ \\
llama3.3\_70b & QoT--NoQoT & $0.40 \pm 2.97$ & $3.40 \pm 2.07$ & $-3.00 \pm 3.46$ \\
\bottomrule
\end{tabular}

\vspace{10pt}

\textbf{(b) QoT $-$ CoT}

\begin{tabular}{l c rrr}
\toprule
\textbf{Model} & \textbf{Setting} & \textbf{$\Delta$API} & \textbf{$\Delta$Comm} & \textbf{$\Delta$FS} \\
\midrule
llama3.1\_8b  & QoT--CoT & $2.00 \pm 1.73$ & $2.40 \pm 3.05$ & $1.20 \pm 2.77$ \\
llama3.1\_70b & QoT--CoT & $5.80 \pm 1.30$ & $6.60 \pm 0.89$ & $3.20 \pm 1.48$ \\
llama3.2\_3b  & QoT--CoT & $3.60 \pm 2.51$ & $1.40 \pm 1.67$ & $1.40 \pm 5.86$ \\
llama3.3\_70b & QoT--CoT & $2.20 \pm 2.28$ & $4.80 \pm 2.17$ & $2.20 \pm 3.90$ \\
\bottomrule
\end{tabular}

\end{table*}

\begin{table*}[t]
\centering
\normalsize
\setlength{\tabcolsep}{8pt} 
\renewcommand{\arraystretch}{1.2} 
\caption{Condensed example of QoT: we pair an 8-step sequential chain with self-questioning themes. The counts (n=*) show how many self-QA prompts we use at each step, and we provide the full list in the artifact.}
\label{tab:qot_example_compact}
\begin{tabular}{c l l}
\toprule
\textbf{Step} & \textbf{Sequential Process Chain (API Design)} & \textbf{Self Q-A focus (examples)} \\
\midrule
1 & User model + registration/login & uniqueness, password security, data schema (n=7) \\
2 & Room model + CRUD & naming constraints, reservation dependency, permission checks (n=8) \\
3 & Reservation lifecycle & double-booking, concurrency, cancellation rules, storage (n=10) \\
4 & Notification module & trigger mechanism, required fields, delivery/logging (n=8) \\
5 & Route architecture & endpoint decomposition, relations, extensibility (n=8) \\
6 & Flask implementation & HTTP methods, error handling, query params (n=10) \\
7 & In-memory storage policy & reset semantics, consistency, limits/cleanup (n=8) \\
8 & Integration into app.py & module boundaries, exception safety, maintainability (n=8) \\
\bottomrule
\end{tabular}
\end{table*}

We operationalize each criterion as a checklist of observable evidence in the generated artifacts (e.g., presence of validation and error paths, clear module boundaries, and explicit controls such as password hashing and access checks). For each task output, we document the evidence used to assign scores and release the rubric and scoring worksheet alongside raw generations. This design enables independent re-scoring and supports transparent auditing of the reported quality gains.

\subsection{Experimental Results}
\begin{figure*}[!t]
    \centering
    \includegraphics[width=0.75\textwidth]{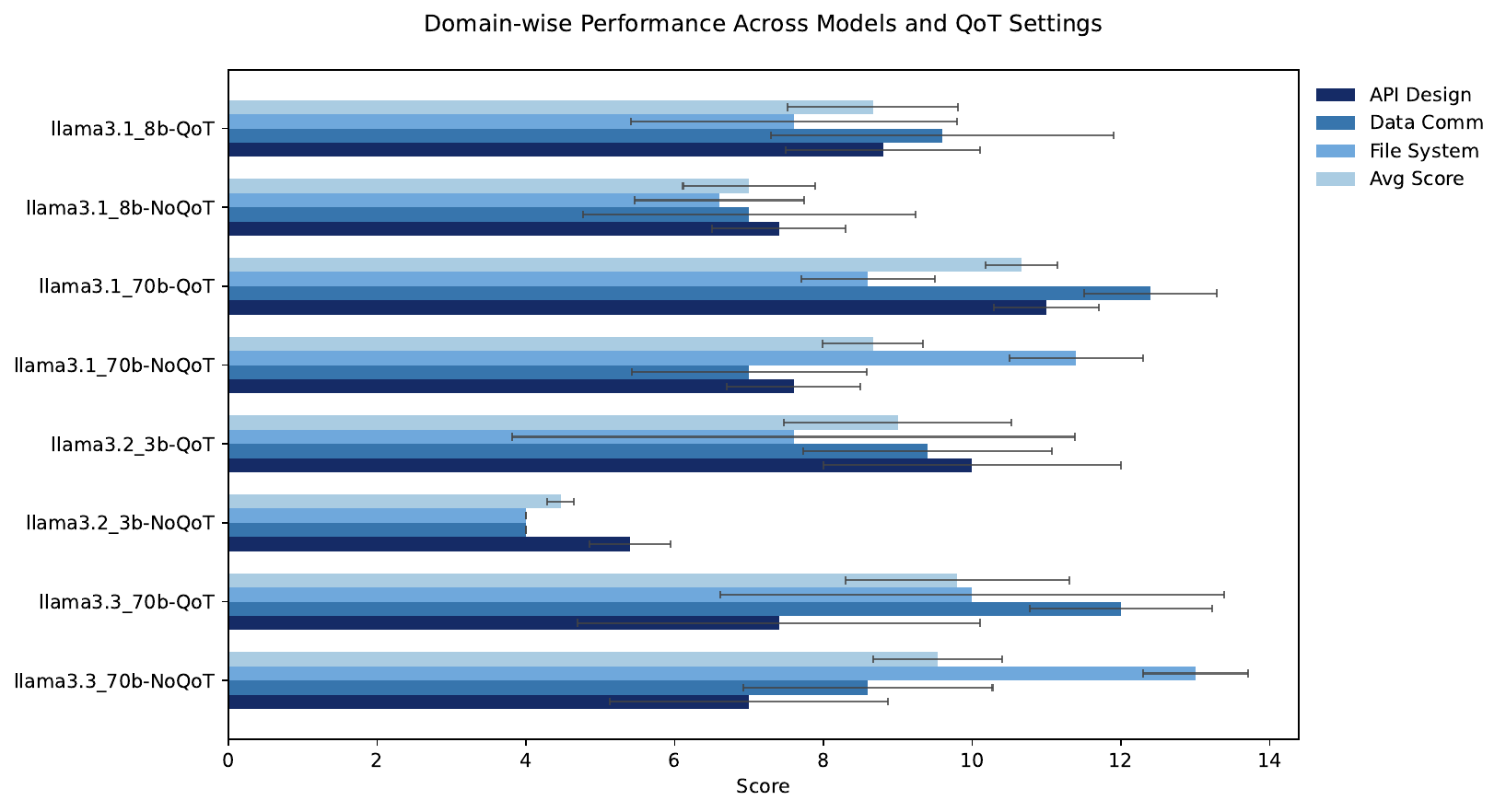}
    \caption{Overall experimental results. QoT generally improves quality scores relative to non-QoT baselines, with model- and domain-dependent effects.}
    \label{fig:results}
\end{figure*}

\begin{figure*}[!t]
\centering

\includegraphics[width=\textwidth]{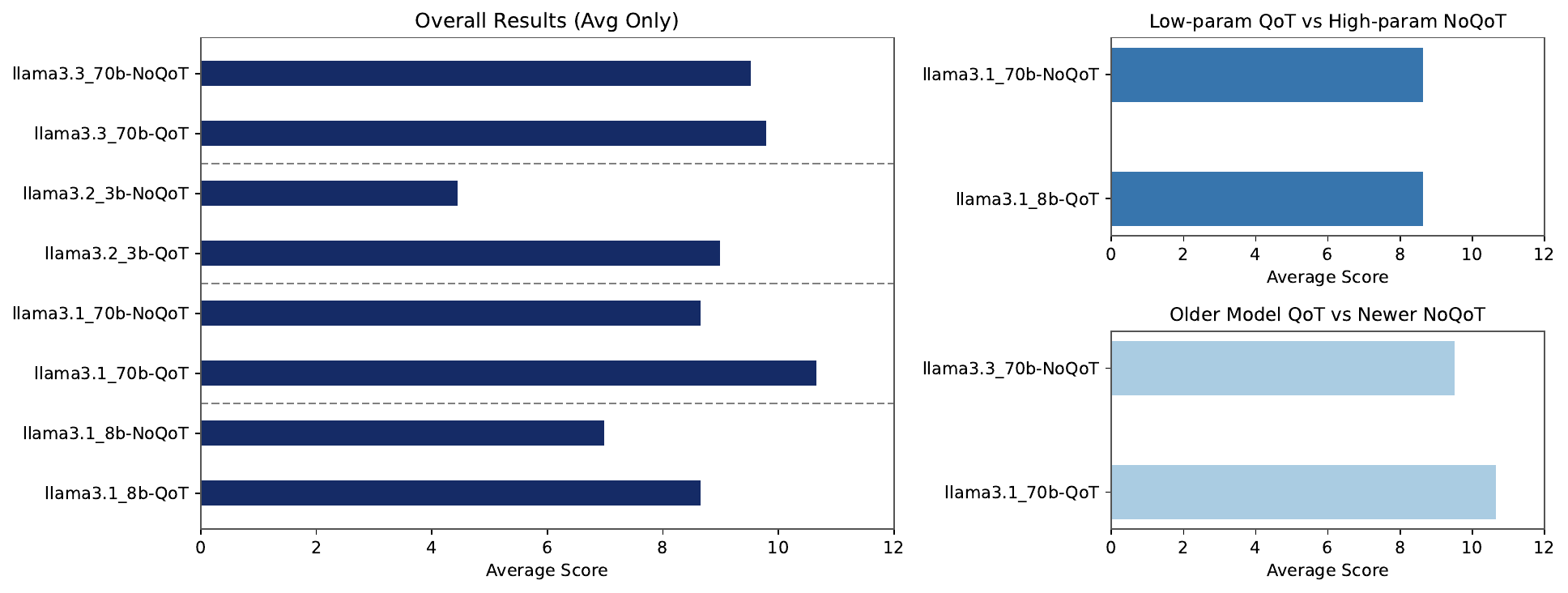}

\vspace{0.5em}

\normalsize
\setlength{\tabcolsep}{8pt}
\renewcommand{\arraystretch}{1.2}

\footnotesize
\setlength{\tabcolsep}{8pt}
\renewcommand{\arraystretch}{1.15}

\begin{tabular}{l c @{\hspace{1.5cm}} l c}
\toprule
\multicolumn{2}{c}{\textbf{QoT vs NoQoT}} & 
\multicolumn{2}{c}{\textbf{Cross-model}} \\
\midrule
\textbf{Model} & \textbf{Imp.} &
\textbf{Comparison} & \textbf{Imp.} \\
\midrule
llama3.3\_70b & 2.80\% &
llama3.1\_8b-QoT vs 3.1\_70b-NoQoT & 0.00\% \\
llama3.2\_3b & 101.49\% &
llama3.1\_70b-QoT vs 3.3\_70b-NoQoT & 11.89\% \\
llama3.1\_70b & 23.08\% & & \\
llama3.1\_8b & 23.81\% & & \\
\bottomrule
\vspace{0.5em}
\end{tabular}

\caption{Detailed experimental results. Domain-wise comparisons across model configurations with and without QoT highlight capacity-dependent gains and occasional trade-offs. The table summarizes percentage improvements observed in key comparisons.}
\label{fig:detailed_results}

\end{figure*}

QoT improves code quality across the three benchmark domains, with the most significant gains observed in modular decomposition, input validation, and security-related controls.
Figure~\ref{fig:results} reports the overall comparison between QoT and non-QoT settings, while Figure~\ref{fig:detailed_results} provides a model-by-model breakdown.

To characterize QoT's effect across model capacities and generations, we analyze three recurring comparisons: (i) \emph{QoT vs.\ no-QoT within the same model}, (ii) \emph{smaller models with QoT vs.\ larger models without QoT}, and (iii) \emph{older models with QoT vs.\ newer models without QoT}.
Rather than summarizing these qualitatively, we quantify the net effect using the change in total quality score ($\Delta$Total = QoT$-$NoQoT) for each domain, shown in Table~\ref{tab:qot_gain}.

Table~\ref{tab:qot_gain} shows that QoT brings consistent improvements compared to the CoT baseline across most models and domains. In particular, in the large model \texttt{llama3.1\_70b}, QoT demonstrates substantial improvements of +5.8 in API design and +6.6 in data communication, and further achieves an improvement of +3.2 in the file system domain. \texttt{Llama3.3\_70b} also shows a similar positive trend where QoT improves performance in all three domains compared to CoT.

For smaller models (\texttt{llama3.1\_8b} and \texttt{llama3.2\_3b}), QoT still provides certain advantages, but the improvement is relatively moderate. They show improvements in domains compared to CoT, such as API design and data communication. However, the variance observed in some settings suggests that the effectiveness of self-questioning may depend on the model’s ability to maintain consistent planning and reasoning across multiple steps. One possible explanation is that smaller models may generate lower-quality intermediate questions during the self-questioning stage. When the reasoning capacity of the model is limited, these intermediate questions may fail to capture key design constraints, which can destabilize the accumulated reasoning context and reduce the effectiveness of the QoT reasoning scaffold.

We also observe a performance degradation for large models in the File System (FS) domain, with \texttt{llama3.1\_70b} and \texttt{llama3.3\_70b} yielding negative score changes ($-2.80$ and $-3.00$, respectively). This phenomenon suggests a potential risk of over-thinking or over-engineering. 

Ultimately, these results highlight a trade-off: while QoT serves as a powerful scaffold for complex tasks, its success requires a delicate balance between the model's multi-step planning capacity and the inherent complexity of the task to avoid both reasoning degradation and over-engineering.

To provide an interpretable view of why QoT improves quality, Table~\ref{tab:qot_example_compact} summarizes a condensed QoT trace on API Design: an 8-step sequential plan paired with stepwise self-questioning themes (e.g., constraints, security, dependency management).
We include the full self-QA transcripts and generated artifacts in the released repository.

\subsection{Findings and Observations}
From the results, we observe three consistent takeaways:
\begin{itemize}
    \item \textbf{Enhanced Modularity and Security:} QoT-enabled models more often generate layered validation, access control, and defensive error handling.
    \item \textbf{Improved Performance for Small Models:} Smaller models gain substantial quality improvements under QoT, reducing dependence on huge models.
    \item \textbf{Consistency Across Domains:} QoT benefits appear across all tested domains, suggesting general applicability to everyday software-engineering tasks.
\end{itemize}

Overall, these results position QoT as a practical, quality-driven reasoning protocol for applied AI, agentic systems, and trustworthy software design.

\section{Discussion}

QoT targets a practical gap that appears when LLM agents are deployed in real engineering and analytics pipelines. In such settings, functional correctness alone is insufficient because the generated system must also satisfy operational constraints, preserve traceability, and support auditing and long-term maintenance. QoT addresses this gap by turning code generation into a criteria-driven workflow that makes dependencies explicit, checks constraints step by step, and accumulates reusable requirements in an evolving reasoning knowledge base. By structuring reasoning around software quality dimensions such as scalability, completeness, modularity, and security, QoT transforms generation from a single-pass synthesis task into a traceable process where intermediate reasoning artifacts reveal how design constraints influence the final implementation. This design aligns well with emerging agentic software engineering workflows that combine planning, tool use, and iterative refinement in applied AI systems.

QoT introduces additional inference steps because it executes a structured self-questioning loop at each stage of the reasoning plan. Let $S$ denote the number of reasoning steps and $Q$ the number of self-questions generated per step. While standard prompting performs a single inference pass with cost $O(T)$, QoT performs a structured loop with approximate complexity $O(S \times Q \times T)$, where $T$ represents token generation cost. In practical deployments, the absolute latency depends on hardware configuration and inference throughput, but the additional overhead mainly arises from generating intermediate questions and answers. In many engineering tasks this cost is acceptable because reducing design errors can significantly lower downstream debugging and maintenance effort. QoT is therefore most suitable for high-stakes scenarios such as backend service design, integration workflows, and compliance-sensitive software components where reliability and maintainability outweigh minimal latency.

Although QoT improves reasoning transparency and process reliability, it does not guarantee correctness or eliminate hallucinations. Trustworthy deployment therefore requires combining structured reasoning with automated verification mechanisms. A promising direction is to integrate QoT with verification pipelines such as static analysis, unit-test synthesis, and policy validation so that generated artifacts are accompanied by machine-checkable evidence. In addition, future work can explore adaptive reasoning strategies that terminate self-questioning when quality gains saturate, as well as segmented reasoning approaches that store intermediate artifacts such as module interfaces, validation rules, and threat models in external memory for long-horizon tasks. These directions could further improve scalability, governance alignment, and trustworthiness in agentic software engineering systems.

\section{Conclusion}

We presented \textit{Questions-of-Thoughts} (QoT), a quality-driven agentic reasoning framework for LLM-assisted software design.
QoT integrates sequential planning, structured self-questioning, and an evolving reasoning knowledge base to guide code generation toward deployable software quality targets such as scalability, completeness, modularity, and security.
Across a curated benchmark covering API design, data communication, and file systems, QoT consistently improves quality scores compared with non-QoT baselines and often enables smaller models to approach the quality level of larger models in single-pass generation settings.

Beyond improving functional correctness, QoT structures the reasoning process around explicit software quality criteria and surfaces design constraints throughout the generation workflow.
This design transforms code generation from a single-shot synthesis task into a traceable, criteria-driven reasoning process that supports inspection, auditing, and iterative improvement of generated artifacts.

As LLMs increasingly power real-world engineering and analytics pipelines, practitioners require reliable and maintainable outputs rather than fluent code alone.
QoT addresses this need by embedding quality-oriented reasoning into the generation process.
Future work will integrate QoT with automated verification pipelines, including static analysis, test synthesis, and policy-based validation, to further strengthen trustworthy agentic software engineering.

\appendix
\section{Reproducibility Notes}
We release prompts, model configurations, scoring rubrics, and evaluation scripts in our open-source repository (\url{https://github.com/knyliu/questions-of-thoughts}). The repository also includes scripts for running the QoT pipeline and reproducing the evaluation tables and figures reported in this paper.

\bibliographystyle{IEEEtran}
\bibliography{bib/references}

\end{document}